# Privacy Preservation Intrusion Detection Technique for SCADA Systems

*Marwa Keshk, Nour Moustafa, Elena Sitnikova, Gideon Creech*
School of Engineering and Information Technology
University of New South Wales at the Australian Defence Force Academy, Canberra, Australia
E-mail: marwa.hassan@student, nour.abdelhameed@student, g.creech, e.sitnikova, @{.adfa.edu.au}

*Abstract—* **Supervisory Control and Data Acquisition (SCADA) systems face the absence of a protection technique that can beat different types of intrusions and protect the data from disclosure while handling this data using other applications, specifically Intrusion Detection System (IDS). The SCADA system can manage the critical infrastructure of industrial control environments. Protecting sensitive information is a difficult task to achieve in reality with the connection of physical and digital systems. Hence, privacy preservation techniques have become effective in order to protect sensitive/private information and to detect malicious activities, but they are not accurate in terms of error detection, sensitivity percentage of data disclosure. In this paper, we propose a new Privacy Preservation Intrusion Detection (PPID) technique based on the correlation coefficient and Expectation Maximisation (EM) clustering mechanisms for selecting important portions of data and recognizing intrusive events. This technique is evaluated on the power system datasets for multiclass attacks to measure its reliability for detecting suspicious activities. The experimental results outperform three techniques in the above terms, showing the efficiency and effectiveness of the proposed technique to be utilized for current SCADA systems.**

*Keywords- SCADA systems, Privacy preservation technique, Intrusion Detection system, EM clustering, Correlation Coefficient,*

I. INTRODUCTION

Supervisory control and data acquisition (SCADA) systems that operate in critical infrastructure are progressively executed with Internet-based protocols and devices for remote control. The involved embedded components' nature and legacy factors add non-trivial and novel security approaches efficiently [1]. In a SCADA system, it is significant that a protection mechanism is embedded to define various types of intrusive activities and protecting sensitive information from disclosure. It monitors and controls critical industrial infrastructure utilities, such as gas, electricity, waste, traffic, water, and railway [1] [2].

The augmented complexity and connectivity of SCADA systems face a large number of cyber security vulnerabilities [2]. In addition, they include embedded devices that allow only limited protection against current sophisticated strategies of intrusive activities, resulting in exposing the principles of Confidentiality, Integrity and Availability (CIA). Practically, malicious or unauthorized access to external sources, using the TCP/IP model, can threaten SCADA systems by misusing communication faults/problems to launch sophisticated attacks which lead to catastrophic failure, denial of service, compromising the safety and stability of power system operations [3].

To defend against threats to a SCADA system and protecting its shared data, an IDS is defined as a technique for monitoring host or network activities to detect possible threats that face them, has become a powerful solution. Data mining and machine learning algorithms are widely used for establishing an effective IDS [4] because of their capability to handle a large volume of historical data, as it is hard for people to deduce primary traffic patterns in such SCADA's data. However, the scarcity of SCADA data is considered as one of the main issues for establishing an efficient IDS for SCADA systems. The privacy preservation of these data requires a new methodology for keeping these data without disclosure. So, it is widely realised nowadays that SCADA data privacy and confidentiality are greatly becoming a key aspect of data sharing and integration [3] [5].

The aim of designing an intelligent IDS for SCADA systems without revealing their data has become one of the big issues in this field due to several reasons [1][2][3]. First and foremost, the use of traditional machine learning and data mining mechanisms for discovering intrusive activities requires a large amount of SCADA data to be correctly learned and validated. This exposes the principles of privacy and confidentiality [2]. If the mechanisms are learned using a small amount of SCADA information, they will poorly detect those malicious activities. Secondly, selecting the important SCADA's information that helps these mechanisms has become an arduous task because the information of SCADA should be carefully analysed to avoid sensitive information from the processing by the IDS techniques. Finally, the way of capturing information from different incompatible layers and protocols of current SCADA systems demands efficient tools and mechanisms to be successfully handled at a real time monitoring and detection [1] [2]. In the literature, privacy preservation data mining techniques have emerged in order to protect sensitive information [1] [3] [4], but the above limitations for SCADA data are still an active area of research.

This paper addresses the above limitations by applying the privacy preservation concept for SCADA systems. A new privacy preservation Intrusion Detection (PPID) technique is proposed using the correlation coefficient technique for selecting the important information without exposing sensitive information of SCADA data. This uses EM clustering algorithm for identifying intrusive observations of SCADA instances. This technique is evaluated on power system datasets [21] for multiclass attacks. The experimental results outperform three techniques in the above terms, showing the efficiency of it to be utilised for current SCADA systems.

## II. BACKGROUND AND RELATED WORK

### A. IDS and its relation to SCADA systems

Intrusion Detection Systems (IDSs) for SCADA systems have become an active domain of research [2][3][4][5]. IDSs are classified into two types: a host-based IDS monitors the events of a host by collecting information about activities which occur in a computer system, while a network-based IDS monitors network traffic to identify remote attacks that take place across that network. IDS methods are categorized into two types: Misuse-based and Anomaly-based. A misuse-based IDS monitors host or network data audits to compare observed behaviours with those on an identified blacklist. Although it offers relatively high detection rates (DRs) and low false positive rates (FPRs), it cannot identify new attacks [3] [4].

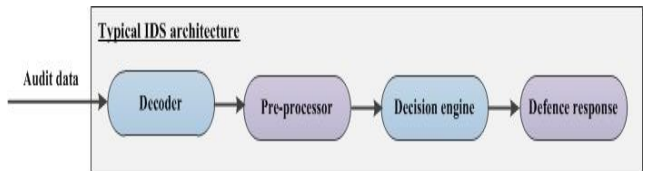

Figure 1. Architecture of typical network IDS [6]

A typical network IDS comprises of four components, namely a packet decoder, a pre-processor, a decision engine sensor and a defence response/alert module [6], as shown in Figure 1 and described as follows.

- The packet decoder acquires portions of raw network traffic using audit data collection tools.
- The pre-processor captures a set of features from the raw audit data, used later in the decision engine sensor.
- The Decision Engine (DE) sensor receives the proposed features by the processor and builds a model that distinguishes attack observations from normal ones. If an attack is detected, the DE requests the defence response to raising an alert.
- The defence response is the process of raising alerts requested by the DE.

Several IDS methods have been proposed for SCADA systems. These are classified into supervised and unsupervised techniques [2] [7] [8]. The supervised methods, such as SVM [8] and Naive Bayes [9], include two main steps. Firstly, the learning step (where classification algorithms aim to analyse the supervised training data and build an inferred model). Secondly, the classification step (where the established model can be utilised to distinguish, detect, and classify malicious and infected data and potential cyber attacks in the future traffic streams). However, one of the major problems with all the supervised techniques is that the step of learning process needs a large size of training observations, which lead to establishing an effective generalised model [2].

It is important to note that the training data should be labelled in advance, which is time-consuming and requires previous knowledge. These Clustering [10] and SOFM–PCA [11] approaches provide some practical advantages over supervised techniques, which can use only unlabelled data. Unsupervised techniques attempt to create clusters in unlabelled traffic data and allocate any future traffic flow to either normal or attack category, according to the nearest cluster. Empirical evaluations demonstrate the higher-purity clusters, found by unsupervised techniques, and detect traffic flows from previously unknown types of applications. However, the unsupervised techniques have a crucial problem in mapping a great number of clusters to real traffic varieties, but they can detect existing and new attacks without labelling processing which is difficult in real time [2] [10] [11].

### B. Privacy Preservation Techniques for sensitive data

Privacy Preservation techniques have emerged as a new direction of research by Aggarwal et al. in 2008 [12] in order to protect retrieving private/sensitive information. The target of these techniques is to considerably reduce unauthorised accessing of sensitive information while running any normal technique to do a particular task, such as IDS, for discovering useful knowledge. Simultaneously, attackers try to retrieve that information that can be executed by normal techniques, breaching the Confidentiality, Integrity and Availability (CIA) of systems. Privacy-preserving mechanisms generally modify the integrity of data to prevent normal techniques from the same knowledge of the altered data as completely and correctly similar to the original data [14].

Numerous privacy-preserving methods for various scenarios of data have been proposed in the past years [2] [13]. The privacy-preserving methods, which modify or transform data to preserve privacy without compromising security, have been roughly classified into three classes, data generalisation methods, data transformation methods and data aggregation methods.

Firstly, in data generalisation methods, the core process is developed based on mapping sensitive attributes to more generalised values. Secondly, in transformation methods, the confidentiality of the data is preserved by transforming the original data into new values, with some random multiplication or by applying projection to the original data, resulting in lower dimensional random space. Finally, in data aggregation methods, the original data is partitioned into a group of small size, and then replaces the private values in each group with the group average. These methods still face the problem of handling different data types effectively [2] [13] and the integration with the IDS method which detects attacks. However, these methods face the challenge of using them for IDS as they provide high false alarm rate of detection, as well as the difficulty of using in SCADA systems that have several layers of data capturing with the linking with the internet services, as discussed below.

*C. SCADA systems*

SCADA (Supervisory Control and Data Acquisition) is defined as the business automation control system central to many modern industries, including, oil, energy, gas, power, transportation and water [2]. Both public-sector providers and private enterprises use SCADA systems, and they can work well in various types of companies because they have the ability to extend from simple configurations to huge, complicated projects. SCADA systems organize manifold software and hardware components, which allow industrial institutions to gather, monitor, and process data, and also, cooperate with control technologies that are connected over Human Machine Interface (HMI) software and record events into a log file [4][15].

Information of SCADA systems is collected from sensors or other manual inputs, and then sent to programmable logic controllers (PLCs) or remote terminal units (RTUs), from which this information is sent to the computers with SCADA software. Current SCADA systems have Ethernet connectivity to enable connecting with the functionalities of networking. Standard protocols include IEC 60870-5-101/104, IEC 61850 and DNP3. These protocols are standardised to operate over the TCP/IP model. Modbus TCP Protocol is commonly supported in devices with Ethernet Connectivity [16]. The main function of SCADA software is to analyse and display the data to help operators and other employees to decrease waste and improve efficiency in the manufacturing procedure, as shown in Figure 2 [5][15].

Some recent studies have focused on designing IDS for SCADA systems, and these studies show the challenges of designing an effective IDS–based SCADA system without disclosing their shared data. Classic encryption techniques, comprising RSA and AES, are not appropriate because we cannot use the encrypted data for analysing this data [17]. Encryption techniques are still the practical solution to prevent disclosing data, during transactions to that data, such as analysis, logging in databases and detection purposes, however, it has to be decrypted first. Meanwhile, processing any of those transactions could be intrusive actions happen; hence it is still a controversial area of research to find methods that provide privacy-preservation for SCADA data [1].

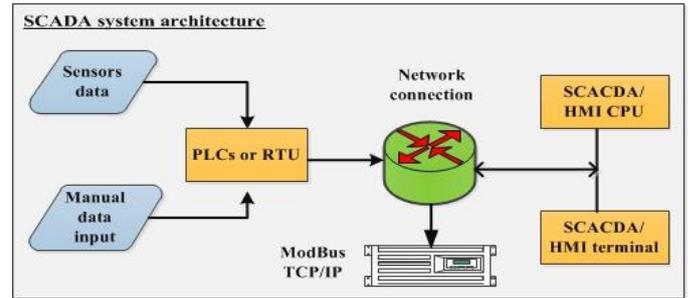

Figure 2. Basic architecture of SCADA systems [9]

## I. RESEARCH METHODOLOGY

This section describes the proposed methodology for building an effective privacy preservation intrusion detection technique for SCADA systems.

*A. Correlation coefficient for selecting important information*

In this subsection, the Pearson's correlation coefficient (PCC) technique is used in order to select the important information of SCADA systems, which is considered as one of the simplest linear correlation techniques for approximating the correlation scores of features [18], with that of two features ($f_1$ and $f_2$)

$$PCC(f_1, f_2) = \frac{cov(f_1, f_2)}{\sigma_{f_1} \cdot \sigma_{f_2}}$$

$$= \frac{\sum_{i=1}^{N}(x_i - M_{f_1})(y_i - M_{f_2})}{\sqrt{\sum_{i=1}^{N}(x_i - M_{f_1})^2} \cdot \sqrt{\sum_{i=1}^{N}(y_i - M_{f_2})^2}} \quad (1)$$

where $cov$ and $\sigma$ are the covariance and the standard deviation of these features, respectively, and $M_{f_1} = 1/N \sum_{i}^{N} x_i$ and $M_{f_2} = N \sum_{i}^{n} y_i$ are the means of $f_1$ and $f_2$, respectively.

The result attained from Equation (1) has to be in a specific range of [-1, 1]; where this result is close to -1, shows a strong correlation in a reverse direction. If it is close to 1, there is a strong correlation in the same direction and, if it is close to 0, there is no correlation among these features, where a positive or negative sign indicates that these features have the same or various tendencies, respectively. The PCC values of features are descendingly ranked to select the important features. If there are a set of features and their correlation matrix is constructed, as shown in Figure 5, the highest correlation feature values produce the significant features.

*B. EM clustering for identifying SCADA attacks*

An EM algorithm [19] is implemented to evaluate the maximum likelihood parameters of a statistical model in several states, such as the one where the equations cannot be explained directly. The EM algorithm repeatedly estimates the unknown model parameters with two main phases: the E and M steps.

Firstly, the E step (i.e., expectation) in which the posterior distribution probabilities of the hidden variables are estimated using the present model parameter values. After that, the objects are slightly allocated to each cluster based on this posterior distribution. Secondly, the M step (maximisation) re-evaluates the parameters of the model with the maximum likelihood rule for estimating the fractional assignment. The EM algorithm is guaranteed to catch a local maximum for the model parameters estimated. The steps of the EM technique are presented in Algorithm 1.

Algorithm 1: the EM algorithm main steps.

**Input:** the dataset (x), the total number of clusters (M), the accepted error threshold for convergence € and the maximum number of iterations

**E-step:** calculate the expectation of the whole data log-likelihood.

$$Q\left(\theta, \theta^T\right) = E\left[\log p\left(x^g, x^m | \theta\right) x^g, \theta^T\right]$$

**M-step:** choice a new parameter estimate that maximizes the Q-function.

$$\theta^{t+1} = \arg\max_\theta Q\left(\theta, \theta^T\right)$$

**Iteration:** increment $t = t+1$; repeat step 2 and 3 until the convergence condition is met.

**Output:** a sequence of parameter estimates $\{\theta^0, \theta^1, ..., \theta^T\}$, which represent the success of the convergence criteria.

### C. Architecture of privacy preservation Intrusion Detection Technique

An effective privacy preservation intrusion detection technique is proposed for preventing the disclosure of sensitive/privacy information and detecting malicious observations of SCADA systems. As shown in Figure 3, the architecture of this technique consists of four steps that demonstrate its ease of application for all SCADA types, in particular, power systems that are used in this study. First and foremost, it is an essential step for building privacy and intrusion detection mechanisms collecting SCADA data in a data source to make it easier while preprocessing and during analysis. As SCADA data is collected from different nodes with diverse protocols that are incompatible when handled by machine learning algorithms, this data has to be processed for execution by those algorithms.

Secondly, selecting portions of important features using the PCC technique prevents disclosing private information of SCADA. This is because that some features are neglected and the most important ones will be used by machine learning techniques. Machine learning techniques demand a large number of features to be successfully learned and validated, but this exposes sensitive information of SCADA systems, so adopting small features can address this issue. This can be measured using the new term of 'sensitivity percentage of data disclosure' that means the rate of the feature selected to the entire number of features used in a data source.

Thirdly, we apply the EM technique for clustering normal and suspicious instances of SCADA data in order to estimate the efficiency of identifying attack activities with a small number of features. Finally, the performance evaluation of the proposed PPID technique is computed in terms of error detection/false positive rate (FPR), sensitivity percentage of data disclosure (i.e., feature percentage). The efficiency of the technique can be achieved when these terms are as low as possible, as discussed in Section V.

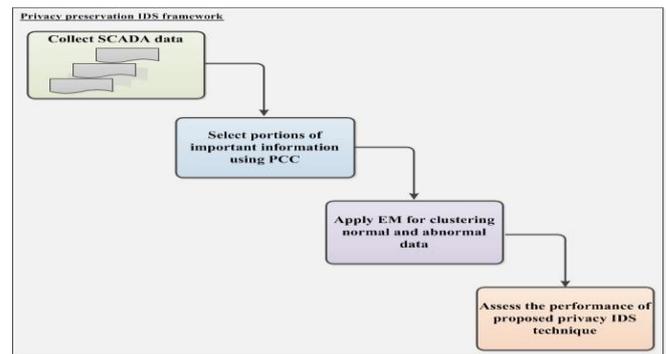

Figure 3. Framework of privacy preservation intrusion detection technique

## II. DATASET FOR EVALUATION

The power system datasets [21] for multiclass attacks are used to evaluate the performance of the proposed privacy preservation intrusion detection technique and are compared with the techniques published in [20]. The multi-class datasets include 37 scenarios, are encompassed into natural events (8), no events (1) and intrusion events (28). The process of establishing this dataset is depicted in Figure 4, which demonstrates the main network components used. Firstly, G1 and G2 are power generators. R1 across R4 are Intelligent Electronic Devices (IEDs) that can turn the breakers on or off. These breakers are tagged BR1 over BR4. There are two lines: line one spans from breaker 1 (BR1) to breaker 2 (BR2) and line two spans from breaker 3 (BR3) to breaker 4 (BR4).

Each IED manages one breaker. R1 controls BR1, R2 controls BR2, where the IEDs apply a distance protection mechanism that trips the breaker on detected faults whether actually valid or fake, as they have no internal validation to identify the difference. Operators can also manually generate instructions to the IEDs R1 over R4 to manually trip the breakers BR1 over BR4. The manual override is utilised while executing maintenance on the lines or other components.

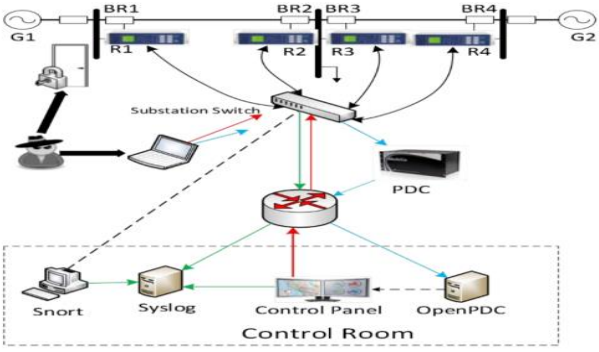
Figure 4. Framework of generating power system datasets [20]

## III. EXPERIMENTAL RESULTS AND DISCUSSION

### A. Evaluation criteria

The accuracy, Detection Rate (DR) and False Positive Rate (FPR) depend on the four terms true positive (TP), true negative (TN), false negative (FN) and false positive (FP). TP denotes the number of actual attack records classified as attacks, TN refers to the number of actual normal records classified as normal, FN is the number of actual attack records classified as normal and finally, FP is the number of actual normal records classified as attacks. These metrics are defined as follows.

- The accuracy is defined as the proportion of all normal and attack records properly classified, that is,

$$Accuracy = \frac{TP+TN}{TP+TN+FP+FN} \quad (2)$$

- The DR is the percentage of precisely detected attack records, that is,

$$Detection\ Rate = \frac{TP}{TP+FN} \quad (3)$$

- The FPR identifies the percentage of incorrectly detected attack records, that equals,

$$False\ Positive = \frac{FP}{FP+TN} \quad (4)$$

### B. Feature Description and selection

In the power system dataset for multiclass attacks, there are 4 synchrophasors that include 29 features for a total of 116 PMU measurements. The features taken from each PMU and their descriptions are shown in Table I.

The PCC technique is employed for ranking these features in a range of [-1, 1], as depicted in Figure 5. Each quarter of these features is applied in order to evaluate the EM technique. The main motive for selecting each portion of these features is to prevent sensitive features from disclosure through the IDS techniques.

Table I. Feature description of power system dataset [20]

| PA1:VH – PA3:VH | Phase A – C Voltage Phase Angle |
|---|---|
| PM1:V – PM3:V | Phase A – C Voltage Magnitude |
| PA4:IH – PA6:IH | Phase A – C Current Phase Angle |
| PM4:I – PM6:I | Phase A – C Current Magnitude |
| PA7:VII – PA9:VII | Pos. – Neg. – Zero Voltage Phase Angle |
| PM7:V – PM9:V | Pos. – Neg. – Zero Voltage Magnitude |
| PA10:VH – PA12:VH | Pos. – Neg. – Zero Current Phase Angle |
| PM10:V – PM12:V | Pos. – Neg. – Zero Current Magnitude |
| F | Frequency for relays |
| DF | Frequency Delta (dF/dt) for relays |
| PA:Z | Apparent Impedance seen by relays |
| PA:ZH | Apparent Impedance Angle seen by relays |
| S | Status Flag for relays |

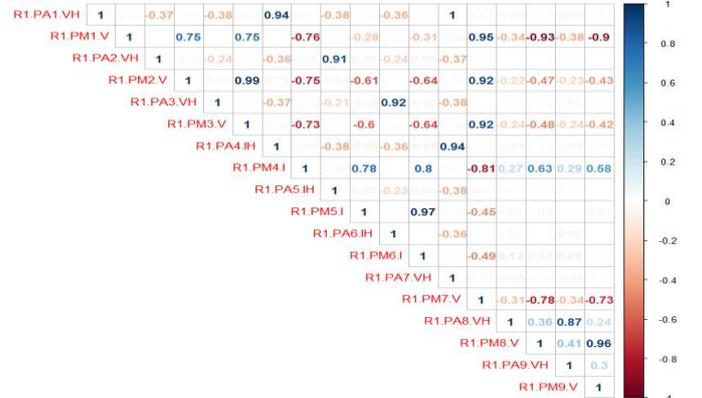
Figure 5. Ranked features using PCC technique

### C. Evaluation using EM technique

The performance evaluation of the EM was conducted in each quarter of these features, revealing the overall DR, accuracy and FPR values demonstrated in Table II. Figure 6 displays the Receiver Operating Characteristics (ROC) curve, which represents the correlation between the DRs and FPRs for the selected features.

Table II. Evaluation of features selection using EM technique

| Feature percentage | DR | Accuracy | FPR |
|---|---|---|---|
| 25% | 66.4% | 70.6% | 32.5% |
| 50% | 75.6% | 76.3% | 25.1% |
| 75% | 82.8% | 83.5% | 17.8% |
| 100% | 88.9% | 90.2% | 11.7% |

It can be observed that the gradual increase of the feature percentages improved the DR and accuracy, whilst dropping the FPR. The DR and accuracy improved from 66.4% to 88.9% and 70.6% and 90.2%, respectively, but the FPR decreased from 32.4% to 11.7%. This reveals that the use of all of the features significantly improves detecting SCADA attacks, however keeping sensitive information secure requires intelligent techniques for detecting malicious events of SCADA systems while using a small number of features, which contains private information.

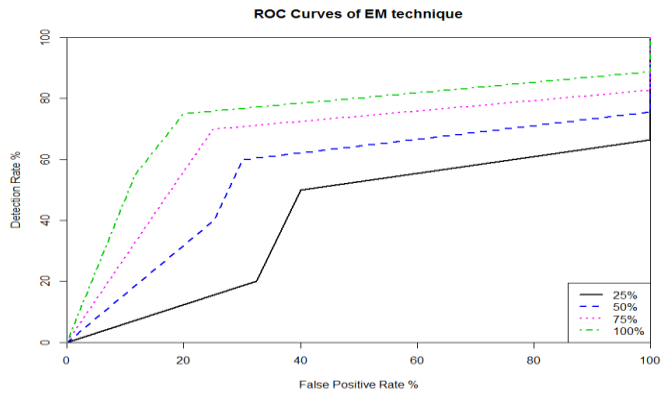

Figure 6. ROC curves of EM technique for feature percentages

*D. Comparison with three techniques*

The results of our technique are compared with the three techniques of Nearest Neighbour, Naïve Bayes, Random Forests published in [20]. As demonstrated in Table III, our technique achieves superiority in terms of DR and FPR compared with these techniques, with the 75% of features.

Table III. Performance comparisons of techniques

| Technique | DR | FPR |
|---|---|---|
| Nearest Neighbour [20] | 55.3% | 44.8% |
| Naïve Bayes [20] | 44.4% | 52.6% |
| Random Forests [20] | 60.5% | 38.4% |
| Our technique | 88.9% | 11.7 % |

Our technique accomplishes better results than the three other techniques, as it clusters normal and abnormal data based on the exact estimation of mean and standard deviation of normal and attack classes, making a clear difference between them. However, the technique is not able to detect the best results with a lower number of features because of the relative similarity between normal and attack data. This can be achieved by using hierarchical clustering techniques to use a small number of features, resulting in disclosing sensitive information of SCADA systems.

## IV. CONCLUSION

This study produces a novel privacy preservation intrusion detection mechanism using the correlation coefficient EM clustering techniques. Important features are selected based on the correlation coefficient technique to select portions of SCADA data with less sensitive information. Then, the EM clustering technique groups SCADA data in order to effectively and efficiently detect abnormal activities. The performance evaluation of this mechanism is compared with three peer techniques using the power system dataset for multiclass attacks, with the superiority of the proposed mechanism for detecting SCADA attacks. The experimental results reveal reducing the number of features that prevent disclosing sensitive information slightly decreases the detection rate of attacks. In future, we will use advanced clustering techniques that significantly improve the detection accuracy.